\documentstyle[11pt,paspconf,psfig,epsf]{article}

\begin{document}

\title{Abundances of  Big Bang elements}
\author{Paolo  Molaro}
\affil{Osservatorio Astronomico di Trieste, Via G. B. Tiepolo 11, I34131, Italy}

\begin{abstract}
In this paper we  review the present status of 
observations of the Big Bang elements D, $^3He$, $^4He$ and $^7Li$
and of their extrapolation to the primordial values. It is shown that, 
within the errors,  the
abundances are consistent with  the predictions 
of the standard Big Bang nucleosynthesis for 
 1 $\le \eta_{10} \le 6$, which corresponds to 
0.004 $ \le \Omega_b h_{0}^2 \le 0.02$.
Narrower consistencies  in $\eta_{10}$ are still possible at $\approx$ 1.7 or $\approx$ 4,  but in this case   some of the observations of D and $^4He$ 
should be incorrect.  
 Finally, it is noted  that {\it extragalactic} Li may  have 
already been  detected in a star 
possibly accreted by the Galaxy.
\end{abstract}


\section{Introduction}

In the standard hot homogeneous Big Bang model with 
3 neutrino flavours the elemental yields of
primordial
nucleosynthesis   
depend only on   the ratio between the
number of baryons and photons  at that  epoch, namely:

\begin{displaymath}
\eta = \frac{n_b}{n_{\gamma}}
\end{displaymath}

The value of $\eta$ is not  predictable by the physics of the early universe.
 It can only be fixed 
from the observations
of the primordial
abundances of the light elements $D,  ^3He, ^4He$  and $^7Li$. 
Unlike   $^4$He, where the yields are sensitive to the  speeding  up of the expansion,  the other three  elements $^3He, D$ and $^7Li$  
 {\it must} show consistency with 
the  same value for $\eta$, whatever  the number of relativistic
 particles at the nucleosynthesis. More details can be found in   Salati (1997)  and Sarkar (1996).  There are four  overall  observables,
 i.e.
 the four light element abundances, and one variable, i. e. $\eta$, 
which make the  Big Bang theory  fully testable. 
This redundancy turns out to be  particularly useful since 
 the determination of the
primordial values of the four elements from the observations is not so straightforward.

The number of photons $n_{\gamma}$ at the epoch of the 
nucleosynthesis 
can be obtained  by scaling back  the present number density of photons of  the
cosmic background radiation (T=2.73 K) since    the stellar
photon production  is  negligible, and, therefore, a simple relation
holds between  $\eta$ and the
the global baryonic density, namely:

\begin{displaymath}
\Omega_b = 0.0037 h^{-2} \eta_{10}
\end{displaymath}
\noindent
where $\eta_{10} = 
10^{10}\eta$ and $h_{0}=$ is the Hubble constant  in units of 100 km s$^{-1}$
Mpc$^{-1}$, with possible values between 0.5 and 1.
 When compared with the luminous and dynamical matter
this value has important bearings for assessing the presence and relative amount of baryonic and non baryonic dark matter respectively.

The  uncertainties in the theoretical yields of SBBN 
come mainly from uncertainties in the nuclear cross-sections. They have been considerably reduced in the recent years after the accurate
determination of   the  neutron half-life decay  of 887($\pm 2$) sec.
The uncertainties  
are  rather small for  $^4$He with $\approx$  0.5 \% at
95 C.L., 
 with  $\approx$  15\% for D and $^3$He, but as large as 
 $\approx$ 50 \% for Li (Krauss and Kernan 1995).
In the following we are using  the  approximate relations to  the theoretical yields over the range of
$\eta_{10}$=1-10 provided by Sarkar
(1996), which are accurate enough  for our purpose.
 These are:
\bigskip
\begin{equation}
Y_p = 0.2462 + 0.01 \ln(\frac{\eta}{5 \cdot 10^{-10}})(\frac{\eta}{5 \cdot 10^{-10}})^{-0.2} \pm 0.0012 
\label{pm:eqn:d1}
\end{equation}

\begin{equation}
\frac{D}{H} = 3.6 \cdot 10^{-5\pm 0.06} (\frac{\eta}{5 \cdot 10^{-10}})^{-1.6}
\label{pm:eqn:d2}
\end{equation}

\begin{equation}
\frac {^3He} {H} = 1.2 \cdot 10^{-5\pm 0.06} (\frac{\eta}{5 \cdot 10^{-10}})^{-0.63} 
\label{pm:eqn:d3}
\end{equation}

\begin{equation}
 \frac {Li} {H} = 1.2 \cdot 10^{-11 \pm 0.2} [ (\frac{\eta}{5 \cdot 10^{-10}})^{-2.38} + 21.7(\frac{\eta}{5 \cdot 10^{-10}})^{2.38}] \label{pm:eqn:d4}
\end{equation}

\bigskip

Determining a primordial abundance is a two-step process. The first one
involves  a measurement of $^{3,4}$He, D and Li in an anvironment, or at least the closest one, 
were pristine
material has been preserved. A measurement of
HI is also required since the significant  abundances are those relative to
 hydrogen. In fact,    the HI determination is often, in particular for deuterium, the most difficult part. Unfortunately, there is no uncontaminated  material available around    and even  the Ly$\alpha$ clouds,  supposed to be made up  of  unprocessed material,
have  revealed  the presence of  metals at the deep Keck observations.
 Probably the sites with the  material closest to the primordial one are
 the   atmospheres of the extreme halo stars,  where the   metallicities 
can be  as low as 0.0001 solar. Therefore, the  second step
  necessarily requires   corrections for
the 15 Gyr or so of  the cosmic stellar pollution.
Recent reviews of the subject are those of Reeves (1994) and Pagel (1995).

\section{The Helium Universal Floor}

 Jakobsen et al. (1994) made a  remarkable  detection of  the 
HeII 304 \AA~ line  in absorption towards the  QSO 0202-003 at
$z_{em}=3.286$ showing    that helium is indeed pervasive in  the universe but, unfortunately, uncertainties in the photoionization
preclude  precise helium measurements.
Helium can be measured by a number of means and in a number of environments
such as  the study of the solar interior,  solar prominences,  He absorption lines in 
hot stars, the position of subdwarfs main sequence, the globular clusters
morphology and the recombination of He lines in planetary nebulae and HII regions.
Among these techniques  the HII regions provide the most  accurate  
helium determination  with
an   accuracy  that can be as small as  $\pm 2\%$. 
 In the HII nebulae hydrogen and helium
emission lines are formed by electronic recombination of H$^+$ and He$^+$. 
For  the abundance determination the necessary physical quantities are the 
electronic temperatures  and densities, that  can be obtained by specific line ratios  
( for instance:  $T_e$  from the  $[OIII]\frac{4363}{(4959+5007)}$ and $n_e$ from the $[SII] \frac{6717}{6731}$ ratios).
Searle and Sargent (1972) first recognized    that the 
extragalactic HII regions  IZw18 and IIZw40  showing  low metal abundances
are the best sites for primordial helium determination.
The extragalactic HII regions or blue compact galaxies (BCG)  are dwarf irregular galaxies undergoing an intense burst of star formation
but characterized by low abundances. The burst is not necessarily the first one
and some of the BCG  show presence of  red stars or Wolf Rayet 
  revealing  previous older
generations of stars which may have contributed to the elemental production and as well as to a small fraction of helium.

 These objects are   those specifically used for 
 cosmological purposes since they 
have   the  virtues  of being
the closest objects to the primordial material and  at the same time of permitting accurate measurements for helium.  So far  about 80 BCG  with a   metallicity range
between $Z_{\sun}/50 < Z < Z_{\sun}/3$ have been studied for helium.
Following Peimbert-Torres and Peimbert (1974)  
the stellar production of helium is given by   the correlation among helium and the "metallicity", namely:
 
\begin{displaymath}
Y = Y_p + Z(\frac{dY}{dZ})
\end{displaymath}
\noindent
where Y stands  for helium, Z   for  oxygen, or  nitrogen as suggested by Pagel
and co-workers,  and the primordial value is obtained by an
extrapolation  at zero metallicity. Since the time scales for oxygen production
are  much
shorter than  for helium an often followed alternative approach is that of taking the mean of the most metal poor 
galaxies.  Strictly speaking this  is just an upper limit to the primordial value.

From  the  first  determination of the mass fraction of primordial helium
of Y$_p$=0.230$\pm$ 0.004  made  by Lequeux et al. (1979)  to
the most  recent one of 0.228 $\pm$ 0.005  by Pagel et al. (1992), the  determinations   have  remained rather stable
around these values  with 
the focus  on the value of  the third decimal place  (see Pagel 1995 for a detail account of the works). 
More recently  Izotov Thuan and Lipovetsky (1994, 1997)
claimed a  primordial helium considerably higher than generally assumed earlier. The two papers are based on the analysis of 10 and 27 new extragalactic HII regions from  the I and II Byurokan objective
prism surveys. The Izotov et al. most  recent value  is:

\begin{displaymath}   
Y_p=0.243\pm  0.003.
\end{displaymath}
\noindent
  The same result is obtained  by using  either  O or N,
and it is also not sensitive to the inclusion of objects with W-R features.
According to the authors,   the $Y_{p}$ value rises up with the use 
of the new HeI recombination coefficients by Smits (1996).
However,  
 as pointed out by
Peimbert 1996 and Olive, Skillman and Steigman (1997),  Smits' coefficients are 
almost  the same of the old ones by  Brocklehurst (1972),
with the exception of the He 7065 \AA~ line, which has been used by Izotov
et al. but not by  previous authors. 
 Olive et al. reanalyzed the whole data set of available extragalactic HII
regions comprehensive of Izotov et al.' ones  for a total of 78
BCG. From the linear regression at zero metallicity they obtained 

\begin{displaymath}
Y_{p}= 0.234 \pm 0.002
\end{displaymath}

 The result does not change    when the sample is reduced to the 62 objects of better data quality, but when the subsample 
of the most metal poor galaxies ([O/H]$<$-1) is  considered, 
the linear regression gives a somewhat lower value Y$_p$=0.230$\pm$0.003. The 
  same value (Y$_p$ = 0.230$\pm$ 0.004) is obtained by averaging  the multiple independent determinations of 
IZw18 which, with [Fe/H]=-1.8, is the most metal poor known among the blue
compact galaxies.
Thus the problem seems to be related to the absence of very metal poor objects   in the Izotov et al.  sample.
 Unless they are giving 
 lower  helium values  for some unknown systematics,   the more likely value for the primordial
helium ranges  between 0.230-0.234.  
In Fig 1 the $Y_p$ abundances are compared with the SBBN predictions for 3 types of neutrinos and  a $\tau_{n}$=887 sec.
The Olive et al. value 
in terms of $\eta_{10}$ gives 1.8$^{+0.9}_{-0.5}$, considering a significance of 2 $\sigma$ both in the observations and in the theoretical uncertainties. By comparison the Izotov et al. value corresponds to a  $\eta_{10} =3.7 ^{+4.1
}_{-1.6}$ at the same value of confidence.

\begin{figure}
\psfig{figure=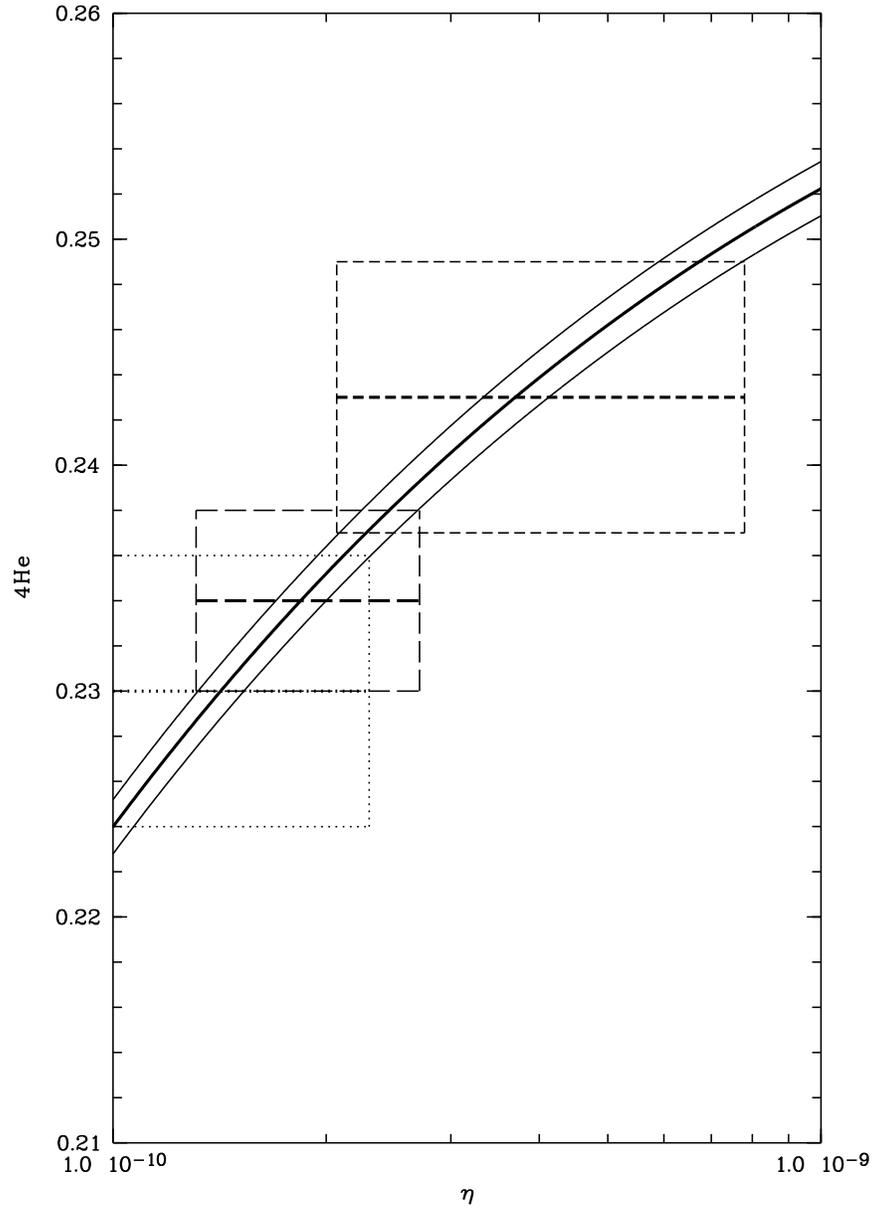,width=12cm,angle=0}
\vspace{1cm}
\caption{SBBN yields for $^4He$ with the observations of Izotov et al. (1997)
and Olive et al. (1997),  errors at 2$\sigma$ of CL.}
\label{pm:fig-1}
\end{figure}

\section{The $^3$He puzzle}

$^3$He is an element presenting many difficulties for its measurement
and interpretation.
The pre-solar system abundance is derived from the non solar component in the meteorites and  is:

\begin{displaymath}
\frac{^3He}{H} = 1.5(\pm 0.3) \cdot 10^{-5}
\end{displaymath}

\noindent
 (Eberhardt 1974).

A new datum has been recently acquired with the solar wind ion
composition spectrometer (SWICS)  on Ulysses spacecraft which measured 
the
$^3He/^4He$ ratio in the interstellar gas entering the solar system
(Gloeckler and Geiss 1996)
along 40 months of integration.  Relative to hydrogen the  value becomes:

\begin{displaymath}
\frac{^3He}{H} =  2.1(^{+0.9}_{-0.8}) \cdot 10^{-5}
\end{displaymath}

 This  measure    refers to the present composition of $^3He$ and it  shows that the $^3He$ abundance has not
changed significantly in the last 4.5 Gyrs.

$^3He$  is also  detected in emission through the 3.46 cm hyperfine transition
of $^3He^+$ in   Galactic HII regions. Data   from a heroic project started
in the early '80s at the 43m Green Bank radiotelescope by Rood and collaborators
 have been obtained
 from about 
14 nebulae (Rood et al. 1995). The derived abundances  are in the range of
 $0.68  \cdot 10^{-5}$  up to 6.03 $  \cdot 10^{-5}$,  probably with
a real dispersion. In addition, the data  suggest   an   anticorrelation with the
galactocentric radius, with the more distant nebulae showing also the largest abundances.

$^3$He is produced in small mass stars (M$\le 2 M_{\sun}$) but it is destroyed 
in more massive stars (Iben 1967).   
All    chemical
evolution models predict a net increase of $^3He$ with time, and  therefore
we should
have  at any time  $^3He_p < ^3He$. 
 But, if we take the value for the S209 nebula, which is the lowest value observed in the Galactic HII
regions, we get  a lower limit on  $\eta_{10} > $ 5, which is inconsistent with 
the bounds coming from the other elements.

According to the review  of Tosi (1996), 
a common feature of all the chemical models  is   an astonishing
 overproduction of $^3He$
by a factor 10 to 40 when compared to the observed one. Furthermore, these models
cannot reproduce the constancy of the element from the solar birth up to
the present time and 
 predict a galactocentric gradient which is opposite
to the  observed one.
Only {\it non standard} models which incorporate some $^3He$ destruction
in low mass stars (Hogan 1995, Charbonnel 1995) 
 can reproduce the observations. 
 However, counterexamples came from the  observations  of 
 $^3He/H$  at the level of 10$^{-3,-4}$ in six PNe with low mass progenitors
in agreement with   the  theoretical predictions of $^3He$ production
in low mass stars (Rood et al. 1995).

In summary  $^3He$ is the most elusive and ambiguous among the quartet of Big Bang elements    both for what concerns the observations and 
 the   understanding of its theoretical behaviour;  therefore, it is of little use for cosmology. Under the assumption that it has not
changed very much we have a  

\bigskip
\centerline{ $^3He_p \approx  2 \cdot 10^{-5}$}
\bigskip
\noindent
which is slightly more  than a guess at  its primordial value.

\begin{figure}
\plotone{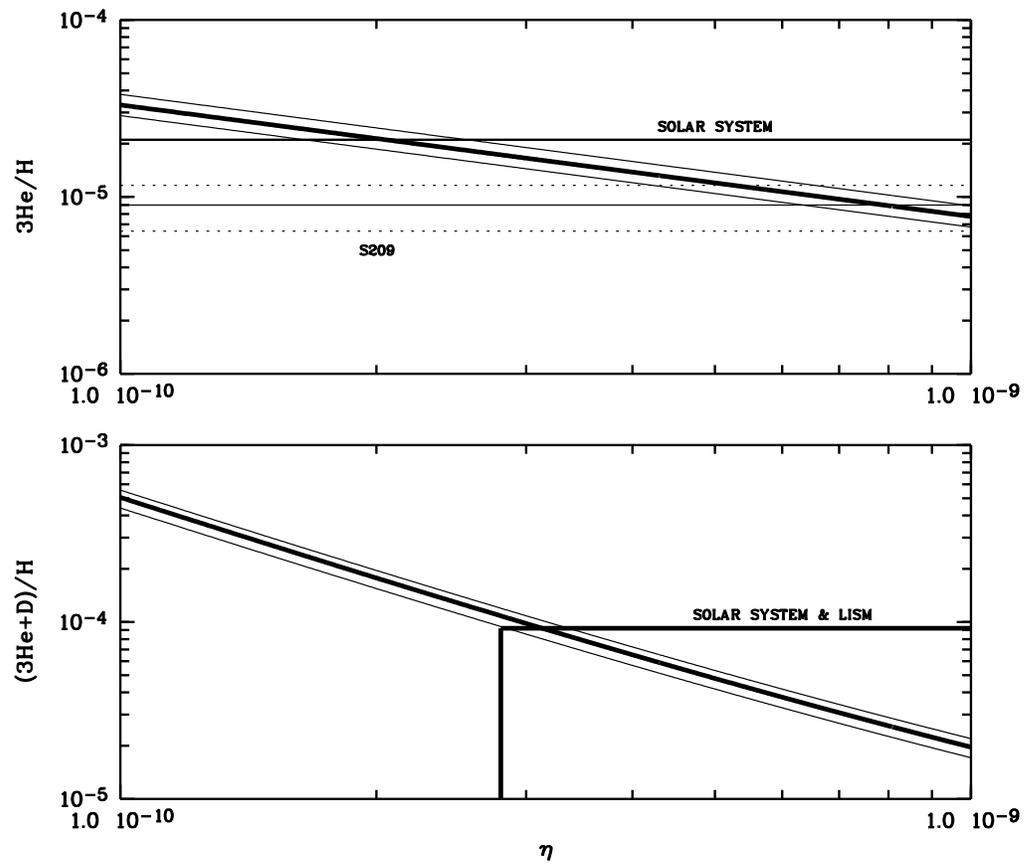}
\caption{Top figure: SBBN $^3He$ yields with 2$\sigma$ errors. Observations are the solar $^3He$
and  S209 nebula from Rood et al. (1995). Botton: The SBBN yields for $(D+^3He)$ and the upper limit for the sum, see text for details.} 
\label{pm:fig-2}
\end{figure}

\section{ The Deuterium controversy}

Primordial nucleosynthesis is the only known source of deuterium, and 
the mere
presence  of deuterium  is an important proof for BBN.
After  all D 
 has been produced in the first three minutes or so,   it is slowly destroyed
in the stellar recycling along the following 15 Gyrs. Correspondingly,
the primordial deuterium   is  higher than  any observed value, i.e.
 $(D/H)_p > (D/H)_o$.


The (pre)-solar system abundance is indirectly deduced by using the fact 
that  when  deuterium is burnt, it is all converted into $^3He$
(Geiss and Reeves, 1972).  By subtracting the present $^3He$ abundance obtained from solar flares from the $^3He$ pre-solar value obtained  from the meteorites,   we obtain  the deuterium abundance    
originally present in the sun prior to deuterium burning:


\begin{displaymath}
\frac{D}{H}=(\frac{^3He}{H_{\sun}}  - \frac{^3He}{H_{pre-\sun}})  = 2.6(\pm 1) \cdot 10^{-5} 
\end{displaymath}

 This is in substantial agreement with the various determinations
in the giant planets of the solar system, all in the range between
2 and 5 $10^{-5}$ (see Griffin et al.  1996, and references therein). 

In the interstellar medium deuterium is measured in absorption from the DI 
Lyman series
 in the  spectra of ultraviolet background sources, 
blushifted by $\approx$ 82 km s$^{-1}$ with respect to
 the
  corresponding HI Lyman serie. 
  A summary  of the best data  from
the Copernicus and IUE satellites towrds hot and relatively distant stars gives
a value of $1.5 \cdot 10^{-5}$,  but variations of a factor 2 
among the different lines of sight are not ruled out (McCullough 1992).
Note that in few cases the D/H ratio was found at a level of $\approx$
5,8 $\cdot  10^{-6}$ towards $\delta$ and $\epsilon$ Ori, $\lambda$ Sco and $\theta$ Car and there is not a clear understanding of such a low values.
 
Historically these determinations of  low deuterium abundance in the interstellar medium and solar system
 provided the first evidence
that $\Omega_{b} \le 0.04h^{-2}$, i.e. that the universe cannot be closed by baryons (Reeves et al. 1973).

High quality data have been recently supplied by  HST.  Measurements of the DI  Ly$\alpha$ have been performed 
 towards the closeby cool stars  Capella, Procyon and $\alpha$ Cen
where the Di and HI Ly$\alpha$ are detected in absorption  on the stellar chromospheric Lyman 
emission (Linsky et al. 1993, 1995, Linsky and Wood 1996). 
Lemoine et al. (1996) measured the D/H towards the featureless continuum
of the hot white dwarfs G191-B2B at a distance of 48 pc.   Three independent clouds are detected along this  line of sight and all of them 
give an  abundance of  D/H $\approx$ 1.3 $\cdot 10^{-5}$.
Towards Capella the interstellar medium is 
remarkably simple with only one component, and  it probably provides
the most  accurate  measure for D/H in the 
local interstellar medium. From  
 Linsky et al. (1995):

\begin{displaymath}
\frac{D}{H} = 1.6(\pm 0.09)_{stat} (^{+0.05}_{-0.10})_{syst} \cdot 10^{-5}
\end{displaymath}

 In general,  difficulties may arise from the modelling of intervening clouds.
 The binary system $\alpha$ Cen 
the system is only 1.34 pc away, which is  the  shortest line of sight one may think,
 and nevertheless   a model with two clouds is required, with a second component contributing to
HI but not to deuterium. The D/H abundance  would be   a factor 2 lower ( i.e. $D/H=0.61 \cdot 10^{-5}$),
 ignoring this  second component, which Linsky and Wood  associate with the compression of the interstellar gas by the solar wind near the heliopause.
 The case of $\alpha$ Cen is a good 
example of how difficult it may be measuring exactly  deuterium within 
 a complex interstellar structure.
 
An important detection is  the long search for   the  92 cm DI hyperfine emission line  reported by
Chengalur et al. (1997). 
The detection is at a significance of  4 $\sigma$ and implies
an    abundance  of

\begin{displaymath}
\frac{D}{H} = 3.9 \pm 1.0 \cdot 10^{-5}
\end{displaymath}

\bigskip
 
\begin{figure}
\psfig{figure=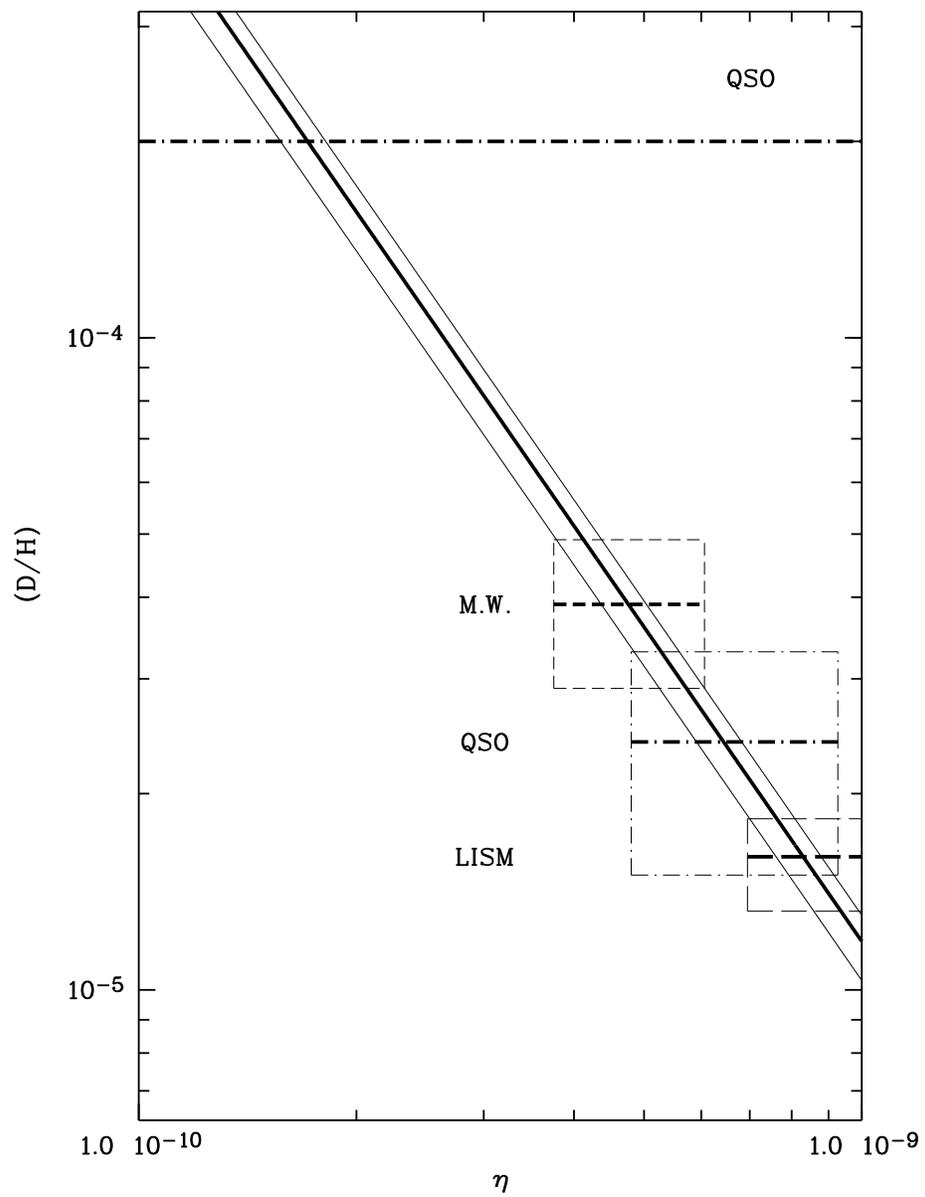,width=12cm,angle=0}
\vspace{1cm}
\caption{SBBN D yields with 2 $\sigma$ errors. Observations are
for the  QSOs, the Galactic anticenter, and the LISM, see text for details} \label{pm:fig-3}
\end{figure}

 The Chengalur  et al. (1997) measure is
 obtained in the direction of the  Galactic anticenter in  a region of the Galaxy that should be less affected by  stellar recycling than the solar neighborhood. This measure   is certainly  the measure of deuterium
closest 
to the primordial value  we may get  in the Galaxy.

The extrapolation to  the primordial value from the measurements
of the abundance in the interstellar medium requires the understanding
of the amount of matter  recycled into stars ({\it astration}).
The ratio between the observed deuterium and the primordial
value $\frac{D}{D_p}$ is the fraction of gas 
that has never been through stars. According to
Tosi's review
the most likely value for this fraction  is 0.5,  whereas  a value of 0.3  seems
a safe  bound. However, the presence of infall  in the Galaxy is an additional variable since it requires pre-knowledge of
the amount of deuterium there.
Less conventional models  can destroy D up to a factor
of 10, without overproducing metals or $^3He$ (Vangioni-Flam
et al. 1994). However,   the problem of $^3He$ overproduction
in models with large D depletion is relaxed if extra  destruction for $^3He$ is required, as discussed  in the previous section.
\subsection{Primordial ($^3$He + D)}

Some of the  problems in  the understanding of  D evolution may be circumvented
by considering  the sum (D+$^3$He) instead of the separated elements. This is 
because any D destruction leads 
to $^3He$ production, some of which survives stellar processing.
Following Yang et al. (1984) at any time the sum

\begin{displaymath}
[\frac{(D+ ^3He)}{H}]p < [\frac{(D+ ^3He)}{H}]_{\sun} + (\frac{1}{f} -1) ^3He_{\sun}
\end{displaymath}
\noindent
where $f$ is the fraction of survival of $^3He$.
Stellar evolution theory  predicts $f >$ 0.25, with a likely value of 0.5.
Thus:

\begin{displaymath}
 [\frac{(D+^3He)}{H}]p < 8.6(\pm 0.3) \cdot 10^{-5}
\end{displaymath}

\noindent
which implies $\eta_{10} \ge 2.5$. This argument was  used   in combination with an upper bound in primordial helium to set a limit  at four for   the number of neutrino flavours (Yang et al. 1984).
However,  most surprisingly, (D + $^3He$) computed at the solar birth has  
 not changed in  comparison with the value of  
 the present epoch, which can be obtained  by combining the data for D and $^3He$
from the  local interstellar medium  
 ($3.6 \cdot 10^{-5}$). This again reflects the poor understanding of the $^3He$ evolution, with the possible presence of unrecognized $^3He$ sinks.
Thus,  the  (D + $^3He$)
 argument does not appear  a very safe one anymore for setting a lower bound to $\eta$. 

\subsection{ Deuterium in high z QSO absorption systems}

 Adams (1976) first suggested  the possibility of
measuring the D/H ratio in high redshift absorption systems. They are 
supposed to be unevolved systems which  offer
the  advantage of a deuterium  abundance 
before a considerable as well as  uncertain,
stellar destruction.   The first positive results, which 
came out only in the 1994, show  that  the measure is not 
straightforward at all. 

Simulations have been presented by Webb et al. (1991) 
and
Jenkins (1996). The most propitious  case seems  to be the detection of  the D Ly$\alpha$
or Ly$\beta$ of a relatively simple, i.e.  one-component
absorption system, with negligible kinematic broadening  along the line of 
sight.  The total column density has to be rather low
to avoid the saturation of the line, so that the best candidates are the
 Limit Lyman System ( $\log N(HI)\le 17.5$).
 A second possibility  suggested by Jenkins
involves the higher members (Ly$\theta$ to Ly$\pi$) of Damped Ly$\alpha$
absorbers,
which are systems with $\log N(HI) \ge 20.6$. Such a possibility 
 is particularly interesting
since these objects are believed to be the progenitors of the present
day spirals, and specific  programs 
are under way.

So far   detections
 for D/H have been claimed for eight systems, but research  is 
developing very fast.
The first was the system  in the QSO Q0014-813 at redshift z=3.32 
(Songaila et al. 1994, Carswell et al.
1994 and Rugers and Hogan 1996).
This system has  $\log N(HI)$=17.3, no metal lines detected which imply 
[C/H], [Si/H] $<$ -3.0, and is composed of at least 7 hydrogen clouds. 
A  deuterium
line  is identified  for  the most  blueward of the seven components which
form the absorption complex, and the   abundance is:

\begin{displaymath}
\frac {D}{H} \approx 2 \cdot  10^{-4}
\end{displaymath}

The crucial question, here and in general for other deuterium identifications
towards QSOs,
is to discriminate
the feature against the possibility of  a Ly$\alpha$ cloud.
For QSO0014-813 the  redshift coincidence is of $\pm$ 5 km s${-1}$, but the argument is not  definitive, in particular if there is some clustering in the system.
Rugers and Hogan (1996) from the reanalysis of the  Keck data used by Songaila et al. claimed that 
the deuterium feature is  made of  two resolved clouds. The measured 
 line broadening is of  $b \approx 8~ km s^{-1}$, 
( $b= (2KT/m)^{\frac{1}{2}} $), which
is rarely found  (less than 2\%) in  Ly$\alpha$ clouds
thus  
making  the  case for a Ly$\alpha$
 interloper less likely.  However,  new Keck data of the same system  by Tytler (1997) are challenging the double nature of the deuterium line.

 Carswell et al. (1996) proposed a tentative detection of deuterium in the
 system at z$_{abs}$ =3.08 towards QSO 0420-388. 
The hydrogen column density of the
system is of about 10$^{18}$ and the metallicity is rather high at
 $\approx$ 0.1 solar. They acknowledge that 
the presence of the deuterium line improves the fit in the Lyman lines but
that  there is not any
compelling evidence for a deuterium line 
and an interloper Ly$\alpha$ can do the same job. 
Carswell et al.  by-pass the uncertainty on the hydrogen column density  by using the
oxygen abundance assumed constant for all 
components of the system and obtain   D/H $\approx$ 2 $\cdot 10^{-4}$.

Wampler et al. (1996) claimed a deuterium detection in a system at
 z=4.672  towards BR 1202-0725. This system  is likely 
dominated  by a single absorption, it shows  the  high ionization features
of CIV and SiIV  and it may have  a rather high abundance with [O/H]$\approx$=0.3.
This would indicate a potentially  embarassing higher  D/H. 
Other tentative but somewhat even more uncertain  detections, 
all giving high D/H  of $\approx 10^{-4}$,  have been reported 
for another system at z=2.79 towards Q0014-813 by Rugers and Hogan (1996b)
and   towards Q0956+122 and GC 0636+680 (cfr. Hogan 1996).

In contrast with  the high values of D/H,   Tytler et al. 
(1996) and  Burles and
Tytler (1996)    found D/H about one order of
magnitute lower  towards Q1937-1009  and in the z=2.54 system towards Q1009+2956. Both systems, which show several similarities,
have  two components  barely resolved 
with the HIRES-Keck resolution  in several metal lines.
For Q1937-1009 the column densities are of 17.94$\pm$ 0.05 and the carbon 
abundances for the two components are  [C/H]=-3 and -2.2. 
For Q1009+2956 the hydrogen
column density is of 17.46 and [C/H]=-2.9 for both the components. The two systems
have a very similar deuterium abundance, with an average of:

\begin{displaymath}
\frac {D}{H} = 2.4 (\pm 0.3)(\pm 0.3_{sys}) \cdot 10^{-5}
\end{displaymath}

 However, Wampler (1996) showed that an alternative model for Q1937-1009
with 3 components may lead to a reduction in the total   hydrogen column density by a factor 3 or even 4,  having as a side effect   weaker damping
wings and a 25\% increase in the flux below $Ly_{22}$, which are both 
consistent with the data. In fact,
a residual flux  below the Lyman break
was  detected in new Keck observations of Q1937-1009 by Songaila et al. (1997). From
an estimation   of  the continuum in that region which takes into account
 the contribution
of the Ly$\alpha$ forest,  the hydrogen  column density is of $N(HI)=5 (\pm 0.1) \cdot 10^{17}$, i.e. about a factor 2 lower than Tytler et al.'s
estimation.
This increases  the deuterium value in this system at $\approx$ 5 $\cdot 10^{-5}$,
and shows that small errors in the (D/H) measured towards QSOs
ar probably unrealistic.
Considering the revised D/H for one of the two Tytler et al. measurements, and  the
determination towards the Galactic
anticenter,  the most likely value for primordial deuterium is around:

\begin{displaymath}
(\frac{D}{H})_{p} \approx 4 \cdot 10^{-5}
\end{displaymath}

This value also reconciles the determination of the local interstellar medium
measurement with the conventional chemical evolution prescriptions.

\begin{figure}
\psfig{figure=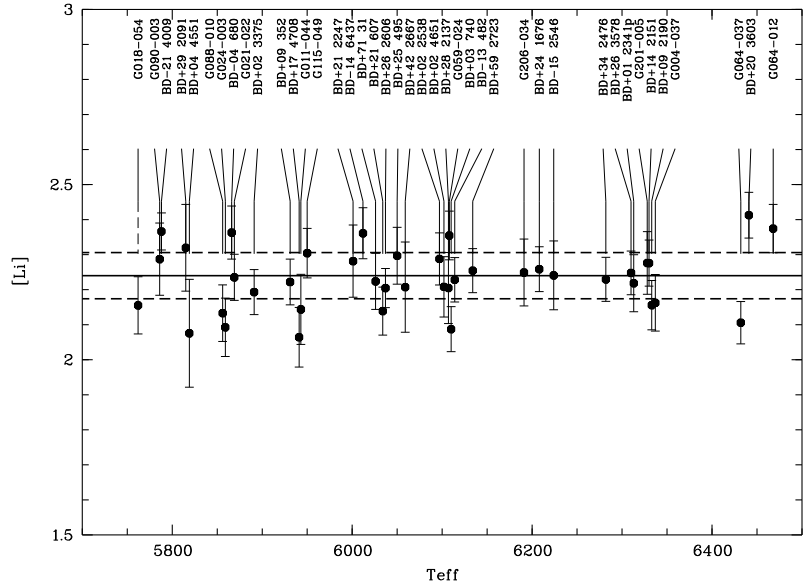,width=12cm,angle=-90}
\vspace{1cm}
\caption{ The Li  {\it plateau} from Bonifacio and Molaro (1997);
$[Li]= \log (Li/H) + 12$.} \label{pm:fig-4}
\end{figure}

\section{Lithium in halo subdwarfs}

Francois and Monique Spite (1982) discovered the presence of  Li and 
its remarkably constant value  in the warm ($T_{eff} \ge  5600 $K) and metal
poor  stars in an amount  very
close to  the minimum of the SBBN yields. Since then 
several
groups have increased  up to about one hundred the number of halo stars studied for lithium, including very metal poor stars with metallicity down
to [Fe/H] $\approx$ -4. 

Recently the results from the field stars have found support in the turn off stars
of globular clusters of NGC 6397 
  (Pasquini and Molaro 1996).   Deliyannis et al. (1995) made 
 Keck observations of few turn-off stars in the globular cluster M92 
and found six stars with about {\it normal} Li except  one   with higher abundance (M92-18).
However, this star has shown some chemical peculiarities and should not 
modify the general result.

The main problem  concerning the cosmological use of lithium   is 
to understand whether  the value measured in  population  II stars
truly represents   the primordial value or  has been depleted from a higher 
value. 
Lithium is a fragile element which is destroyed in the  stellar interiors by (p,$\alpha$)
reactions at $\approx 2.6 \cdot 10^{6}$ K. Thus it survives on  the surfaces
of  F and G stars but
 is depleted in later type   stars with deep convection zones. In addition, 
 other processes, not clearly
identified, should be present depleting further Li and producing the    large scatter of about  3 orders of magnitude, which is observed  in the Li abundances of stars with solar composition.
At variance with  population I,    the Li abundances in   population II  stars are 
closely gathered, with the natural inference  that analogous depletion  mechanisms are
not active  in these stars. 
 The different extent of the convection zones in
the two types of stars is probably responsible  for the different behaviours.
In halo stars the decrease in the 
opacity  creates  a shallower
and more superficial  convection zone, preserving  Li from nuclear burning.

Standard stellar models predict no destruction 
 but   various mechanisms such as
 diffusion, rotational mixing and stellar winds have been investigated as 
a way to deplete Li still  preserving a  plateau
shape (Vauclair and Charbonnel 1995).
Gravitational settling or thermal diffusion operate in very stable atmospheres,  while in the case of rotational mixing the exchange of  material between the surface and the interior is produced by the loss of angular momentum
 which creates  a rotational instability 
inside the star. According to  Vauclair and Charbonnel (1995)   these
 processes 
can deplete the original Li content by a factor 2 or 3, but they
 have some distinct  features such as  a downturn at the
 hot edge of the plateau, a small anticorrelation of Li abundance with
the effective temperature  of the star and a  dispersion of the order of
at least 0.3 dex. 
Stellar mass loss produces 
a readjusting of the stellar structure which is able to produce Li
diluition when Li-free layers enter into the photosphere.
A mass loss higher than  $10^{-13} M_{\sun} yr^{-1}$   starts  bringing up Li depleted layers while    higher
values ($>> 10^{-12}$) lead to complete Li destruction.
The sun has a mass loss of $10^{-14} M_{\sun} yr^{-1}$ and  since no direct observations of winds are possible in population II stars,  this  proposal remains rather speculative. With  stellar winds  of $\approx 10^{-12.5} M_{\sun} yr^{-1}$ a positive slope is predicted  
in the plateau as well as  a considerable dispersion,
unless they are  strictly identical in all the stars.

As a consequence  of these claims the question of the real  flatness  in
the Spite plateau becames  rather important.  Trends of Li-Teff and Li-[Fe/H]
and of intrinsic dispersion have been claimed by some authors
(Deliyannis et al. 1993, Thorburn 1994, Ryan et al. 1995), supporting
 the presence of some Li depletion. On the other hand  the plateau has been found really flat, without any intrinsic dispersion or trend with temperature and metallicity in the analysis of
Molaro et al. (1995), Spite et al. (1996) and Bonifacio and Molaro (1997).
 The different 
results are probably due to  the
different ways to derive the stellar effective temperature for the stars, which is a 
rather critical point for Li abundances. 
 From an accurate reanalysis of 41 stars in the plateau (T$_{eff}$$>$ 5700 and [Fe/H]$<$ -1.5), with available   
 good T$_{eff}$ determinations obtained from
the Infrared Flux Method, Bonifacio and Molaro (1997) have obtained a very small dispersion  of 
$\pm$0.088 around the mean, which is 
of the same order of the observational accuracy. 
The observations are shown in Fig 4.  No trends are found either with the effective temperature or
metallicity, and there is no evidence for a downturn for the Li abundances of the warmer stars.
These results argue against any kind of depletion predicted by
diffusion, rotational mixing or stellar winds, so  that  there are
no real observational features  which may support the case for Li depletion.
 On the contrary, also the Li measurements in tidally locked binaries
and 
 the detection of the more fragile $^6Li$ isotope
in the atmosphere of few halo stars suggest the absence of  Li depletion.

Averaging over all the measurements and after correction for small Non-LTE 
effects  the primordial Li   becomes:

\begin{displaymath}
(\frac{Li}{H})_{p} = 1.73 (\pm 0.05_{stat}) (\pm 0.11_{sys}) \cdot 10^{-10}
\end{displaymath}

where the small statistical error is the error of the mean, and the systematic error, which is   dominant,
follows from  the uncertainty in the
zero point of  the T$_{eff}$ scale of   cool stars (Bonifacio and Molaro 1997). 
 Our ultimate ability to 
model  stellar atmospheres is the only area left in which 
other  systematic errors can be hidden.

In Fig. 5 the   SBBN  theoretical  Li yields together with the
primordial Li from Bonifacio and Molaro (1997) are shown .
The Li Pop II value  corresponds to  
two possible values for $\eta$:
$ \eta_{10} 
= 1.7^{+0.6}_{-0.3}$ 
or $ \eta_{10} =4.0^{+0.9}_{-1.0} $,. 
The minimum  theoretical
Li yield is left out  when one considers $1\sigma$ errors 
in the theoretical predictions and $1\sigma$ errors,
plus the full systematic error,
in the Li abundances.
However, considering  the $2\sigma$ errors, 
the minimum of the Li curve
is allowed thus  
 considerably increasing the allowed $\eta$ range to $1.2 < \eta_{10} < 5.5$ .

\section { Already detected "extragalactic" Li?}

Among the Big Bang elements only D and $^4He$ have been observed in extragalactic objects. Most of the   $^4He$ 
measures  come from the local universe
at redshifts typically of 0.01 to 0.1 at most, and 
deuterium  is observed up to a redshift of 3 or even 4. 
The difficulties  of observations makes  $^3He$   a definitely  Galactic 
element. The same applies to Li, 
 since solar type stars in external galaxies
are out of reach even  for the 10m class telescopes. Only upper limits for 
the interstellar Li towards the SN1987A in the LMC were  obtained when the supernova was as bright as  V=4 mag. 
However, Li  has been   detected in Galactic stars, which might possibly   
have been born  in  
other galaxies. 

Preston et al. (1994) in their HK objective prism survey of metal poor stars 
identified a population
of stars which they called the blue metal poor main sequence stars (BMP). 
This population is composed by  hot and metal poor objects that should have already evolved from the main sequence
if coeval with  the halo stars. 
The space density for the BMP stars is about one order of magnitude larger than that of
blue stragglers in globular cluster, thus suggesting that field BS are a minor component
of this population. Moreover, the kinematical properties of the BMP
are intermediate between those of halo and thick-disk populations.
The Preston et al.'s suggestions is that   the BMP population has been accreted from  a low luminosity
satellite of the Milky Way in the recent past.

One of these stars is CS 22873-139, which has a  remarkably low metallicity
 [Fe/H]=-3.1. This object is
a spectroscopic binary with a period of 19.16 days, and  Preston (1994)
was  able to 
derive an  upper limit to the age for the system at $<$ 8 Gyr, which 
again is  inconsistent
with a halo origin.   Li has been already measured  in CS 22873-139 
by Thorburn (1994) who   derived a value of (Li/H)=1.9 $\cdot 10^{-10}$ , i.e.  at the canonical
 halo  value. Under the assumption that the  object has been truly accreted by our Galaxy, 
this may be considered the first {\it extragalactic} Li
detection giving support to the universality of the Li observed in
the population  II of the Galactic halo. 
 
\begin{figure}
\psfig{figure=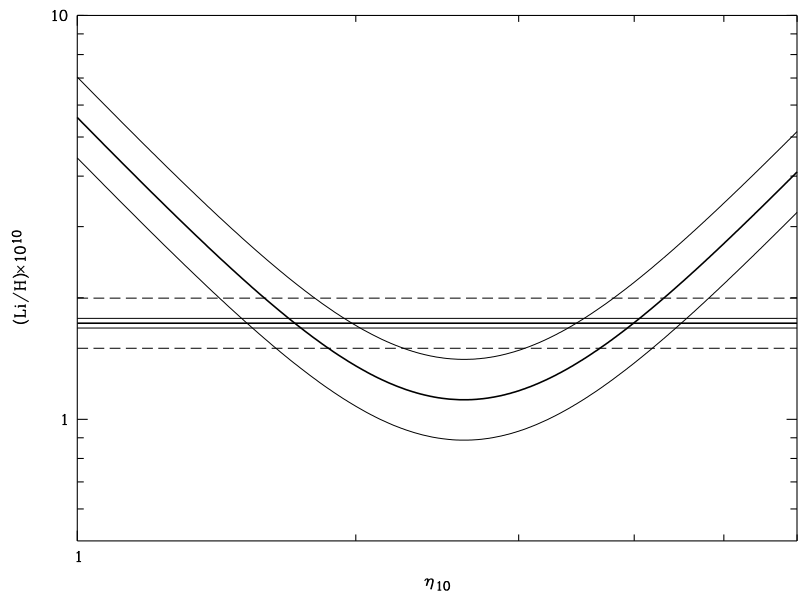,width=12cm,angle=-90}
\vspace{1cm}
\caption{SBBN Li yields with 1$\sigma$ error and $Li_{p}$ from the observations of Bonifacio and Molaro (1997) with 1 $\sigma_{stat}$  + 1 $\sigma_{syst}$.} \label{pm:fig-5}
\end{figure}

\section{ Crisis or concordance}

 Whether or not   concordance  exists between  the observations of the light element 
abundances and  the standard Big Bang  nucleosynthesis
  depends on  the
errors one is willing  to associate to the different measurements.
It is evident that most of the different determinations for a specific 
element are in conflict 
with one another  by a factor
that exceeds the claimed errors.

For deuterium  both  high and low D/H  measures towards QSOs  are clearly
 conflicting.
 The high value may be affected by Ly$\alpha$ interlopers, while
the low value may adopt a too simple cloud model and/or underestimate
the hydrogen column density.
On the other hand,  if they were  both accurate, then the two classes of
objects may have had very different chemical histories. 
It must be realized that 
the real nature and chemical evolution of the systems originating the QSO absorption systems 
are not clearly understood, and 
what we  know today is only that they are probably the  
progenitors of the present day galaxies.
 It has also been pointed out that the total mass covering the QSO image
is rather small ($\le 1 M_{\sun}$),  and local effects as those from 
deuterium-free winds
of massive stars  may have depleted deuterium
significantly along the line of sight. 
Deuterium is the only light element of the quartet  observed on cosmological scales, but  the  possibility  that the different D/H abundances reflect
inhomogeneities in the baryon number at the epoch of nucleosynthesis
on very large scales has been found    inconsistent with  the CRB isotrophy (Copi et al. 1996).     

Concerning Li,  there is no dispute about its  abundance  among the  observers,
and the agreement among the measurements carried out  by
different authors with different theoretical atmospheric models is excellent. 
The controversial issue is the possible depletion, but despite several 
claims there is no observational   support for any depletion. It is a very unfortunate circumstance that
the theoretical uncertainties associated with
the SBBN yields are much  larger than the observational  ones.
 
Since intrinsic variations in the primordial abundances are 
cosmologically unlikely,  unrecognized   
 systematics
are   probably affecting the  elemental measurements.
Adopting the most conservative   approach   to  take the different values 
as an indication of the systematics errors involved in the  measurements, 
all  the  abundances 
 agree with each other for 

\begin{displaymath}
 1 \le \eta_{10} \le 6
\end{displaymath}

   This consistency although not particularly tight 
has to be regarded as a   success of the Big Bang theory which 
is able to predict the right absolute 
abundances for all the four primordial
elements, which differ from one another  by something as nine orders of magnitude. This corresponds to:

\begin{displaymath}
0.004 \le \Omega_b h_{0}^2 \le 0.02
\end{displaymath}

\begin{figure}
\psfig{figure=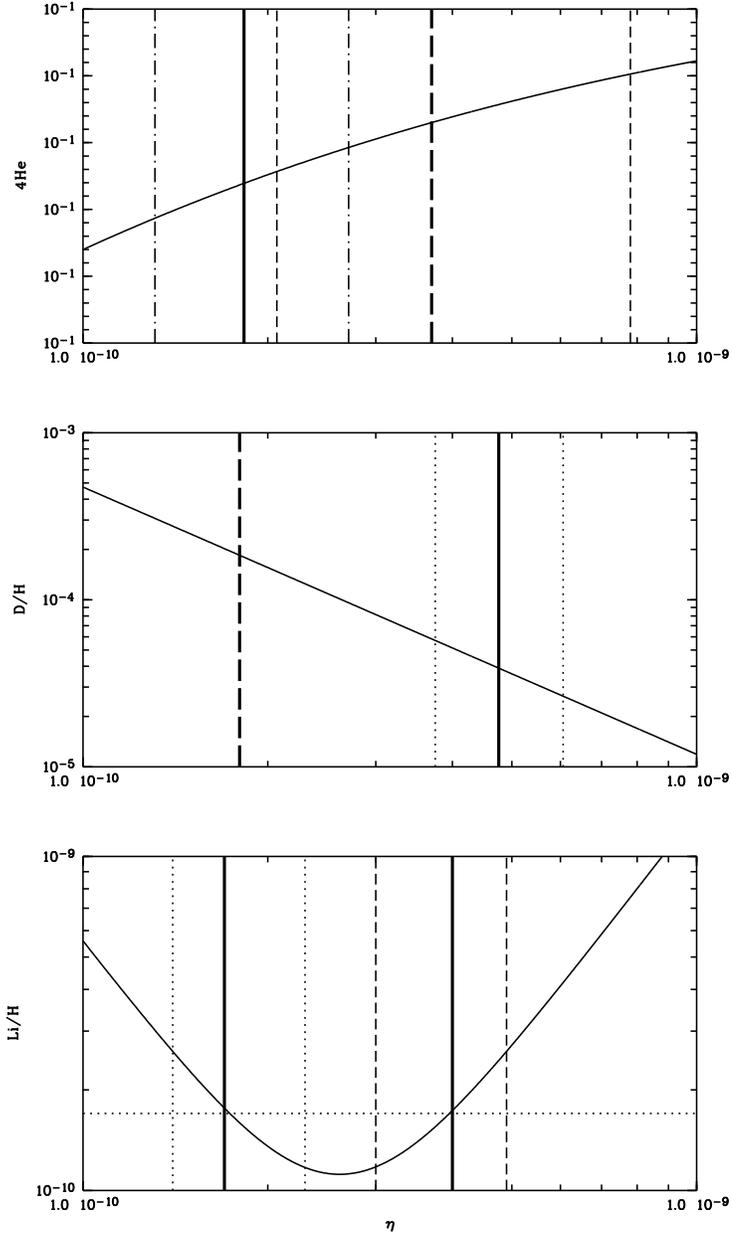,width=12cm,angle=0}
\vspace{1cm}
\caption{ Comparison of theory and observation. $Y_p$ is from Olive et al. (1997)
and Izotov et al. (1997); D is from Chengalur et al. (1997), considered at 1 $\sigma$, and
Li from Bonifacio and Molaro (1997); {\it first choice}  measurements
are shown by continuous lines.} \label{pm:fig-6}
\end{figure}

As it is possible to see in Fig. 6, where the ranges in $\eta$
for  the most likely abundance are shown, 
there may be room  for   two narrower  ranges for  $\eta_{10}$, 
where concordance
is realized by particular choices for  D and $Y_{p}$. 
The two narrower bands are centered roughly in correspondence 
of  the two intersections of the primordial Li value taken
 from the Pop II observations  at face value,
  and  the   valley of the SBBN Li yields.

One narrow  band  is centered at $\eta_{10}$ $\approx$ 1.7.
This is in  excellent
agreement with the   high deuterium suggested by
several measurements in QSO absorption systems and it is also in perfect
agreement with
a  primordial helium at $Y_{p}$=2.34 as derived by Olive et al. (1997)
from the analysis of the {\it whole} data sample of 
extragalactic HII regions.
Two basic problems  affect this consistency. One is the low deuterium observed   by Tytler et al.,
which holds  even if we take the     value revised  by Songaila et al. (1997)
of $\approx$ $4 \cdot 10^{-5}$. The second one is the  difficulty of the
conventional    chemical evolution models to explain an original D/H abundance at a level
of $10^{-4}$ starting from the value of the interstellar medium, which is one order of magnitude
lower.

A slightly higher band  centered at   $\eta_{10}$ $\approx$  4
shows  the concordance   of   deuterium at 
$4 \cdot 10^{-5}$ as it is measured by Chengalur et al. (1997)
from the anticenter region of the
Milky Way. This is about the same value shown by the distant quasar systems of Tytler et al., once
we take into account the new hydrogen column density of
Songaila et al. (1997), but  the high value of D/H observed in some QSO
requires some contamination from Ly$\alpha$.
On this value of  $\eta_{10}$  there is  also the concordance of the Izotov et al. (1997)
determination of helium at $Y_{p}=0.243$, which is also shown in Fig 6.
But  this would imply  that 
the most metal poor extragalactic HII regions lead to a
systematic  underestimation of primordial helium by about 
0.014 in mass.

However, it is rather unsatisfactory that 
{\it first choice}  measures for helium at $Y_{p}$=0.234 and  deuterium  
at (D/H) $\approx 4 \cdot 10^{-5}$  denotes some friction 
among them since they do not strictly match the same value for
$\eta_{10}$ as they are expected to do.
 An alternate solution suggested by Hata et al. (1997) would be to consider
a lower primordial production of helium, which may be achieved by
relaxing the  hypothesis of 3  relativistic neutrinos 
 as assumed in the standard model.

\acknowledgments

I wish to  warmly thank   Timothy  Beers, Gary  Steigman, Francois Spite and   Piercarlo Bonifacio for their valuable suggestions and discussions on the 
the topic of the light
elements.

\end{document}